\documentclass[aoas,nameyear,seceqn,dvips]{arximspdf}
\usepackage{graphics}

\doi{10.1214/09-AOAS312E}
\referstodoi{10.1214/09-AOAS312}
\volume{3}
\issue{4}
\pubyear{2009}
\firstpage{1285}
\lastpage{1294}

\makeatletter

\newtheorem{theorem}{Theorem}[section]

\makeatother

\begin{document}
\begin{frontmatter}

\title{Discussion of: Brownian distance covariance}
\pdftitle{Discussion on Brownian distance covariance by G. J. Szekely
and M. L. Rizzo}
\runtitle{Discussion}

\begin{aug}
\author[A]{\fnms{Arthur} \snm{Gretton}\thanksref{t2}\ead[label=e1]{arthur.gretton@gmail.com}\corref{}},
\author[B]{\fnms{Kenji} \snm{Fukumizu}\ead[label=e2]{fukumizu@ism.ac.jp}}
\and
\author[C]{\fnms{Bharath K.} \snm{Sriperumbudur}\thanksref{t3}\ead[label=e3]{bharathsv@ucsd.edu}}
\thankstext{t2}{Supported by Grants DARPA IPTO FA8750-09-1-0141, ONR
MURI N000140710747, and ARO MURI W911NF0810242. }
\thankstext{t3}{Supported by Max Planck Institute (MPI) for Biological
Cybernetics, NSF
Grant DMS-MSPA 0625409, the Fair Isaac Corporation, and the University
of California MICRO program.}
\runauthor{A. Gretton, K. Fukumizu and B. K. Sriperumbudur}
\affiliation{Carnegie Mellon University, MPI for Biological Cybernetics,
The Institute of Statistical Mathematics, Department of Electrical and
MPI for Biological Cybernetics and Computer Engineering, UCSD}
\address[A]{A. Gretton\\ Machine Learning Department, CMU\\
5000 Forbes Ave \\
Pittsburgh, Pennsylvania 15213\\ USA\\
\printead{e1}} 
\address[B]{K. Fukumizu\\ The Institute of Statistical Mathematics \\
4-6-7 Minami-Azabu, Minato-ku \\
Tokyo 106-8569\\ Japan\\
\printead{e2}}
\address[C]{B. K. Sriperumbudur\\ Department of Electrical\\\quad and Computer
Engineering\\
University of California, San Diego \\
La Jolla, California 92093-0407\\ USA\\
\printead{e3}}
\end{aug}



%
\begin{keyword}
\kwd{Independence testing}
\kwd{Brownian distance covariance}
\kwd{covariance operator}
\kwd{kernel methods}
\kwd{reproducing kernel Hilbert space}.
\end{keyword}

\end{frontmatter}

\section{Introduction}

A dependence statistic, the Brownian Distance Covariance, has been proposed
for use in dependence measurement and independence testing: we refer to
this contribution henceforth as SR [we also note the
earlier work on this topic of \citet{SzeRizBak07}]. Some advantages of the
authors' approach are that the random variables $X$ and $Y$ being
tested may have arbitrary dimension $\mathbb{R}^p$ and $\mathbb
{R}^q$, respectively;
and the test is consistent against all alternatives subject to the
conditions $\mathbf{E}\|X\|_p<\infty$ and
$\mathbf{E}\|X\|_q<\infty$.

In our discussion we review and compare against a number of related
dependence measures that have appeared in the statistics and machine
learning literature.
We begin with distances of the form of SR, equation (2.2), most notably
the work of \citet{Feuerverger93}; \citet{Kankainen95};\break \citet{KanUsh98}; \citet{Ushakov99}, which
we describe in Section \ref{sec:characteristic}: these measures have
been formulated only for the case $p=q=1$, however. In Section \ref
{sec:RKHS} we turn to more recent dependence measures which are
computed between mappings of the probability distributions $\mathbf{P}_x$,
$\mathbf{P}
_y$, and $\mathbf{P}_{xy}$ of $X$, $Y$, and $(X,Y)$, respectively, to high
dimensional feature spaces: specifically, reproducing kernel Hilbert
spaces (RKHSs). The RKHS dependence statistics may be based on the
distance [\citet{SmoGreSonSch07},\break Section 2.3], covariance
[\citeauthor{GreBouSmoSch05bb} (\citeyear{GreHerSmoBouetal05,GreBouSmoSch05bb,GreFukTeoSonetal08})], or\break
correlation [\citet{DauNki98}; \citet{BacJor02};\break \citet{FukBacGre07}; \citet{FukGreSunSch08}]
between the feature mappings, and make smoothness assumptions which can
improve the power of the tests over approaches relying on distances
between the unmapped variables. When the RKHSs are characteristic
[\citet{FukGreSunSch08}; \citet{SriGreFukLanetal08}], meaning that the feature mapping
from the space of probability measures to the RKHS is injective, the
kernel-based tests are consistent for all probability measures
generating $(X,Y)$.

RKHS-based tests apply on spaces $\mathbb{R}^p\times\mathbb{R}^q$
for arbitrary $p$
and $q$. In fact, kernel independence tests are applicable on a still
broader range of (possibly non-Euclidean) domains, which can include
strings [\citet{LesEskWesNob02}], graphs [\citet{GaeFlaWro03}], \ and  groups
[\citet{FukSriGreSch09}], making the kernel approach very general.
In Section \ref{sec:experiments} we provide an empirical comparison
between the approach of SR and the kernel statistic of
\citeauthor{GreBouSmoSch05bb} (\citeyear{GreBouSmoSch05bb,GreFukTeoSonetal08}) on an independence testing benchmark.


\section{Characteristic function-based dependence measures}\label{sec:characteristic}

We begin with a brief review of characteristic function-based independence
measures related to the statistic $\mathcal{V}_n^2({\bf X},{\bf Y})$ in
SR, equation (2.8); see also \citet{Ushakov99}, Section 3.7.

\citet{Feuerverger93} proposes two statistics for independence testing,
in the case where $X$ and $Y$ are univariate. The first,
described by \citeauthor{Feuerverger93} [(\citeyear{Feuerverger93}), Section~4], is
\[
T_{n}:=\int\int\frac{|\Gamma
'_{n}(s,t)|^{2}}{(1-e^{-s^{2}})(1-e^{-t^2})}W(s,t)\,ds\,dt,
\]
where $W(s,t)$ is a weight function,
\[
\Gamma'_{n}(s,t):=f_{\widetilde{X}\widetilde
{Y}}^{n}(s,t)-f_{\widetilde
{X}}^{n}(s)f_{\widetilde{Y}}^{n}(t),
\]
and $f_{\widetilde{X}\widetilde{Y}}^{n}$, $f_{\widetilde{X}}^{n}$, and
$f_{\widetilde{Y}}^{n}$ denote
the empirical characteristic functions (in accordance with the notation
of SR), however, these take as their argument the approximate normal
scores of the sample points,
%
\begin{equation}\label{eq:rankScore}
\widetilde{X}_{i}:=\Phi^{-1}\biggl(\frac{\mathrm
{rank}(X_{i})-3/8}{n+1/4}\biggr).
\end{equation}
With an appropriate choice of weight function, Feuerverger
obtains the statistic
\begin{eqnarray*}
T_{n}^{(1)} & = & \frac{\pi^{2}}{n^{2}}\sum_{j,k}|\widetilde
{X}_{j}-\widetilde{X}_{k}||\widetilde{Y}_{j}-\widetilde
{Y}_{k}|-\frac
{2\pi^{2}}{n^{3}}\sum_{j,q,r}|\widetilde{X}_{j}-\widetilde
{X}_{q}||\widetilde{Y}_{j}-\widetilde{Y}_{r}|
\\
&&{} +\frac{\pi^{2}}{n^{4}}\sum_{j,k,q,r}|\widetilde
{X}_{j}-\widetilde
{X}_{k}||\widetilde{Y}_{q}-\widetilde{Y}_{r}|,
\end{eqnarray*}
where the summation indices denote all $r$-tuples drawn with replacement
from the set $\{1,\ldots,n\}$, and $r$ is the number of indices
of the sum. This statistic takes a form similar to the statistic
$\mathcal{V}_{n}^{2}(\mathbf{X},\mathbf{Y})$
in SR, equation (2.8), the main differences being the restriction to the
univariate case, use of the 1-norm, and transformation (\ref{eq:rankScore}).
A second statistic, described by \citeauthor{Feuerverger93} [(\citeyear{Feuerverger93}), Section 5],
is written
%
\begin{equation}\label{eq:charStat}
T_{n}:=\int\int|\Gamma_{n}(s,t)|^{2}W(s,t)\,ds\,dt,
\end{equation}
where the term $\Gamma_{n}(s,t)$ now simply denotes the difference
between the joint characteristic function and the product of the marginals
[in other words, the statistic is identical to that in SR, equation (2.1)].
Feuerverger remarks that, for certain choices of $W(s,t)$, the resulting
statistic resembles that of \citet{Rosenblatt75}, being the $\ell_{2}$
distance between the kernel density estimate of the joint distribution
and that of the product of the marginals. As an illustration, \citeauthor{Kankainen95} [(\citeyear{Kankainen95}), page~54],
makes this link explicit, employing a Gaussian weight function to
obtain the statistic
%
\begin{equation}\label{eq:hsic}
T_{n}^{(2)} = \frac{1}{n^{2}}\sum_{j,k}k_{jk}l_{jk}-\frac
{2}{n^{3}}\sum
_{j,q,r}k_{jq}l_{jr}+\frac{1}{n^{4}}\sum_{j,k,qr}k_{jk}l_{qr},
\end{equation}
where
%
\begin{equation}\label{eq:kernels}
k_{jk}:=\exp\biggl(\frac{-\Vert X_{j}-X_{k}\Vert^{2}}{2\sigma
_{x}^{2}}\biggr)\quad
\mbox{and}\quad l_{qr}:=\exp\biggl(\frac{-\Vert Y_{q}-Y_{r}\Vert
^{2}}{2\sigma_{x}^{2}}\biggr).
\end{equation}
One can readily see that this involves transforming the distances
of $\mathcal{V}_{n}^{2}(\mathbf{X},\mathbf{Y})$ in~SR, equation (2.8),
by passing them through a Gaussian distortion: this replaces the finite
expected norm condition required by SR with a weaker requirement.

A further difference of \citet{Kankainen95} with respect to \citet
{Feuerverger93}
is that Kankainen generalizes to the problem of testing mutual
independence, although the variables themselves remain univariate.
Kankainen further enforces scale and location invariance by studentizing
each variable. Finally, despite their superficial resemblance, a
number of important differences nonetheless exist
between the statistic in (\ref{eq:charStat}) and that of \citet{Rosenblatt75}.
Most crucially, the kernel bandwidth is kept fixed for the
characteristic function-based test, rather than
decreasing as $n$ rises (a decreasing bandwidth is needed to ensure
consistency of the kernel density estimates),
resulting in very different forms for the null distribution; and
there are more restrictive conditions on the Rosenblatt--Parzen test statistic
[\citet{Rosenblatt75}, conditions a.1--a.4].
These issues are discussed further by \citeauthor{Feuerverger93} [(\citeyear{Feuerverger93}), Section 5],
and \citeauthor{Kankainen95} [(\citeyear{Kankainen95}), Section 5.4]. An empirical
comparison of the null distributions resulting from fixed
vs decreasing bandwidth is provided by \citet{GreGyo08}.


\section{RKHS-based dependence measures}\label{sec:RKHS}

We now present a class of dependence measures
(henceforth \textit{kernel dependence measures})
based on
mappings of the random variables to reproducing kernel
Hilbert spaces,
which encode features of interest for these
variables.
We first use Bochner's theorem to demonstrate
that a subclass of kernel dependence measures is equivalent to SR,
equation (2.2), under appropriate conditions on the weight function.
Next, we give an interpretation in terms of covariances
between feature space mappings, from which we may generalize to broader
classes of kernel dependence measures, including correlations and
estimates of the mean square contingency.


\subsection{Kernel dependence measures via Bochner's theorem}

We describe a dependence measure
introduced by \citeauthor{GreBouSmoSch05bb}
(\citeyear{GreBouSmoSch05bb,GreFukTeoSonetal08})
which constitutes the kernel statistic most closely resembling the
characteristic function-based statistic of SR, equation (2.2).
The present derivation follows \citet{SmoGreSonSch07}, Section 2.3.
We begin with some
necessary terminology and definitions.
Let $z:=(x,y) \in\mathbb{R}^{(p+q)}$, and
$\mathcal{H}$ be an RKHS
with the continuous feature mapping $\theta(z)\in\mathcal{H}$
for each $z\in\mathbb{R}^{(p+q)}$, such that the inner product between
the features is given by the positive definite kernel function
$h(z,z'):=\langle\theta(z),\theta(z')\rangle_{\mathcal{H}}$.
We remark that we never need deal with the feature representations
$\theta(z)$ explicitly
(indeed, these may be infinite dimensional): rather, we express
our statistic entirely in terms of the kernel function, which is the inner
product between two such mappings. If we restrict ourselves to kernels
that can be written in terms of the difference of their arguments,
$h(z,z') = \lambda(z-z')$, the following theorem applies [\citet{Wendland05},
Theorem~6.6].

\begin{theorem}[(Bochner)]\label{Theorem:Bochner}
A continuous function $\lambda\dvtx \mathbb{R}^{(p+q)}\rightarrow\mathbb
{R}$ is positive
definite if and only if it is the Fourier transform of a finite
nonnegative Borel measure $W(u)\,du$ on $\mathbb{R}^{(p+q)}$, that is,
%
\begin{equation}\label{Eq:Bochner}
\lambda(z)=\int_{\mathbb{R}^{(p+q)}}e^{-iz^T u} W(u)\,du,\qquad  z\in
\mathbb{R}^{(p+q)}.
\end{equation}
\end{theorem}

Let us consider the following distance between the joint distribution
$\mathbf{P}:=\mathbf{P}_{xy}$
and the product of the marginals, $\mathbf{Q}:=\mathbf{P}_x \mathbf{P}_y$:
\[
H = \int| f_P(u) - f_Q(u) |^2 W(u) \, du,
\]
where $f_P$ and $f_Q$ are the characteristic functions for $\mathbf
{P}$ and
$\mathbf{Q}$, respectively.
Assuming further that we can decompose $\lambda(z-z')=k(x-x')l(y-y')$
(on which more below), we can rewrite $H$ as
\begin{eqnarray*}
H & =& \int\biggl\{ \int e^{iz^Tu}\,d\mathbf{P}(z) - \int
e^{iz^Tu}\,d\mathbf{Q}(z)
\biggr\}
\\
&&{}\times\biggl\{ \int e^{-iz'^Tu}\, d\mathbf{P}(z') - \int e^{-iz'^Tu}\,d\mathbf
{Q}(z')\biggr\}
W(u)\,du
\\
&=& \int\biggl\{ \int\int e^{i(z-z')^Tu}\, d\mathbf{P}(z)\,d\mathbf{P}(z')
-\int\int e^{i(z-z')^Tu} \,d\mathbf{P}(z)\,d\mathbf{Q}(z')
\\
&&{}\hspace*{14pt}
-\int\int e^{i(z-z')^Tu}\, d\mathbf{Q}(z)\,d\mathbf{P}(z')
+\int\int e^{i(z-z')^Tu}\, d\mathbf{Q}(z)\,d\mathbf{Q}(z')\biggr\}
W(u)\,du
\\
& =& \int\int\lambda(z-z') \,d\mathbf{P}(z)\,d\mathbf{P}(z')
-\int\int\lambda(z-z') \,d\mathbf{P}(z)\,d\mathbf{Q}(z')
 \\
&&{}-\int\int\lambda(z-z') \,d\mathbf{Q}(z)\,d\mathbf{P}(z')
+\int\int\lambda(z-z')\, d\mathbf{Q}(z)\,d\mathbf{Q}(z')
\\
& =&
\mathbf{E}\{k(X-X')l(Y-Y')\}
+\mathbf{E}\{k(X-X')\} \mathbf{E}\{l(Y-Y')\}
 \\
&&{}-2\mathbf{E}\bigl\{\mathbf{E}\{k(X-X')|X\} \mathbf{E}\{k(Y-Y')|Y\}
\bigr\}.
\end{eqnarray*}
We call $H$ the Hilbert--Schmidt
independence criterion (HSIC).
The
test statistic in (\ref{eq:hsic}) is then interpreted as a biased
empirical estimate of $H$ [an unbiased estimate would replace the
$V$-statistics with $U$-statistics; see \citet{GreFukTeoSonetal08}].
We remark at this point that the weight function $1/(|t|_p^{1+p}|s|_q^{1+q})$
is not integrable, hence, Bochner's theorem does not apply for this
choice of $W(u)$.
Thus, interpreting the statistic in SR, equation (2.6), as a kernel
statistic is not
straightforward.

\subsection{Kernel dependence measures via covariance operators}
We now obtain HSIC via a different argument, based on the covariance
between feature mappings of the variables: we then generalize
this to correlation-based dependence measures, with reference
to the statistic $\mathcal{R}^2(X,Y)$ of SR.
Our brief review draws heavily on the overview of
\citet{GreGyo09b}, Section 4.
Let $\mathcal{F}$ be an RKHS on $\mathbb{R}^p$ with feature map $\phi
(X)$ and kernel
$k(X,X'):=\langle\phi(X),\phi(X')\rangle_{\mathcal{F}}$, and
$\mathcal{G}$ be a
second RKHS on $\mathbb{R}^{q}$ with
kernel $l(\cdot,\cdot)$ and feature map $\psi(y)$.
Following \citet{Baker73}; \citet{FukBacJor04}; \citet{GreHerSmoBouetal05}; \citet{FukBacJor09},
the cross-covariance operator
$C_{xy} \dvtx \mathcal{G}\rightarrow\mathcal{F}$
for the measure $\mathbf{P}_{xy}$
is defined such that,
for all $f\in\mathcal{F}$ and $g\in\mathcal{G}$,
\begin{eqnarray*}
\langle f,C_{xy}g\rangle_{\mathcal{F}}
 =
\mathbf{E}\bigl(
[f(X)-\mathbf{E}(f(X))]
[g(Y)-\mathbf{E}(g(Y))]\bigr).
\end{eqnarray*}
The cross-covariance operator can be thought of as a generalization of
a cross-covariance matrix between the (potentially infinite
dimensional) feature mappings $\phi(x)$ and $\psi(y)$.

To see how this operator may be used to test independence, we recall
the following characterization of independence [see, e.g.,
\citet{JacPro00}, Theorem~10.1e]:

\begin{theorem}
\label{th:indep} The random variables $X$
and $Y$ are independent if and only if $\mathrm{cov}(f(X),g(Y))=0$
for any pair $(f,g)$ of bounded, continuous functions.
\end{theorem}

\setcounter{footnote}{2}

While the bounded continuous functions are too rich a class to permit
the construction of a covariance-based test statistic on a sample,
\citet{FukGreSunSch08}; \citet{SriGreFukLanetal08} show that when $\widetilde
{\mathcal{F}}$ is the unit ball in a \textit{characteristic}\footnote
{The reader
is referred to [\citet{FukGreSunSch08}; \citet{SriGreFukLanetal08}] for conditions
under which an RKHS is characteristic. We note here that the Gaussian
kernel on $\mathbb{R}^p$ has this property, and provide further discussion
below.} RKHS
$\mathcal{F}$, and $\widetilde{\mathcal{G}}$ the unit ball in a
characteristic
RKHS $\mathcal{G}$,
then
\[
\sup_{f\in\widetilde{\mathcal{F}}, g\in\widetilde{\mathcal{G}}}
\mathbf{E}\bigl(
[f(X)-\mathbf{E}(f(X))]
[g(Y)-\mathbf{E}(g(Y))]\bigr) = 0
\quad\iff
\quad
\mathbf{P}_{xy}=\mathbf{P}_x \mathbf{P}_y.
\]
In other words, the spectral norm of the covariance operator $C_{xy}$ between
characteristic RKHSs is zero only at independence, and is an
independence statistic [\citet{GreHerSmoBouetal05}].
Rather than the spectral norm, \citet{GreBouSmoSch05bb} propose to use
the squared Hilbert--Schmidt
norm (the sum of the squared singular values), which has a population
expression identical to
HSIC, defined earlier.
The RKHS norm implies a smoothness penalty on the functions $f$ and $g$
\citeauthor{SchSmo02} [(\citeyear{SchSmo02}), Chapter 4], resulting in $O_p(n^{-1/2})$ convergence
of the finite sample estimate: interestingly, this rate does \textit{not}
depend on the dimensions $p$ and $q$ of $X$ and $Y$, respectively.
Following \citeauthor{Serfling80} [(\citeyear{Serfling80}), Chapter 5], the asymptotic distribution of
the statistic under the alternative hypothesis $\mathcal{H}_1$ of dependence
is Gaussian, and the distribution under the null hypothesis $\mathcal{H}_0$
of independence is an infinite weighted sum of independent $\chi^2$
random variables; see [\citet{GreFukTeoSonetal08}] for details.


As long as
$k$ and $l$ are characteristic kernels,
then $ \mathrm{H}(\mathbf{P}_{xy};\mathcal{F},\mathcal{G}) = 0$ iff
$X$ and $Y$ are
independent.
The Gaussian and Laplace
kernels are characteristic on $\mathbb{R}^{p}$
[\citet{FukGreSunSch08}], and universal kernels
[as defined by \citeauthor{Steinwart01b}, (\citeyear{Steinwart01b})]
are characteristic on compact domains [\citet{GreBouSmoSch05bb}, Theorem~6].
\citet{SriGreFukLanetal08} provide a simple necessary and sufficient
condition for a translation invariant kernel to be characteristic on
$\mathbb{R}^p$: the Fourier spectrum of the kernel must be supported
on the
entire domain.
Note that characteristic kernels need
not be functions of the distance between points:
an example is the kernel
\[
k(x,x')=\exp( x^T x' / \sigma)
\]
from \citeauthor{Steinwart01b} [(\citeyear{Steinwart01b}), Section 3, Example 1], which is characteristic
on compact subsets of $\mathbb{R}^p$ since it is universal.
Finally, an appropriate choice of
kernels allows testing of dependence in non-Euclidean settings, such as
distributions on groups, graphs, and strings [see, for
instance, \citet{GreFukTeoSonetal08}, who described independence testing between text fragments in
English and French, where
the null hypothesis was rejected when the French extracts were translations
from the English].

Interestingly, the first RKHS-based independence measures were based
on the canonical correlation, rather than the covariance: in this
respect, they
more strongly resemble the statistic $\mathcal{R}_n^2$ of SR.
\citet{DauNki98} propose
the canonical correlation between variables in a spline-based
RKHS as a dependence measure, using projection on a finite basis to
regularize: this dependence measure follows the suggestion of \citet{Renyi59},
but with a more restrictive pair of function classes used to compute
the correlation (rather than the set of all square integrable functions).
The variables are assumed in this case to be univariate. Likewise,
\citet{BacJor02}
use the canonical correlation between RKHS feature mappings as a
measure of
dependence between pairs of random variables. Bach and Jordan employ
a different regularization
strategy, however, which is a roughness penalty on the canonical correlates.
For an appropriate
rate of decay of this regularization with increasing sample size,
the empirical estimate of the canonical correlation converges in
probability [\citet{LeuMoySil93}; \citet{FukBacGre07}].
Finally, \citet{FukGreSunSch08} provide a consistent RKHS-based
estimate of the mean-square contingency, which is also based on the
feature space correlation. This final
independence measure is asymptotically independent of the kernel choice.
When used as a statistic in an independence test, this last statistic
was found
empirically to have power superior to the HSIC-based test.


\section{Experiments}\label{sec:experiments}

\begin{figure*}

\includegraphics{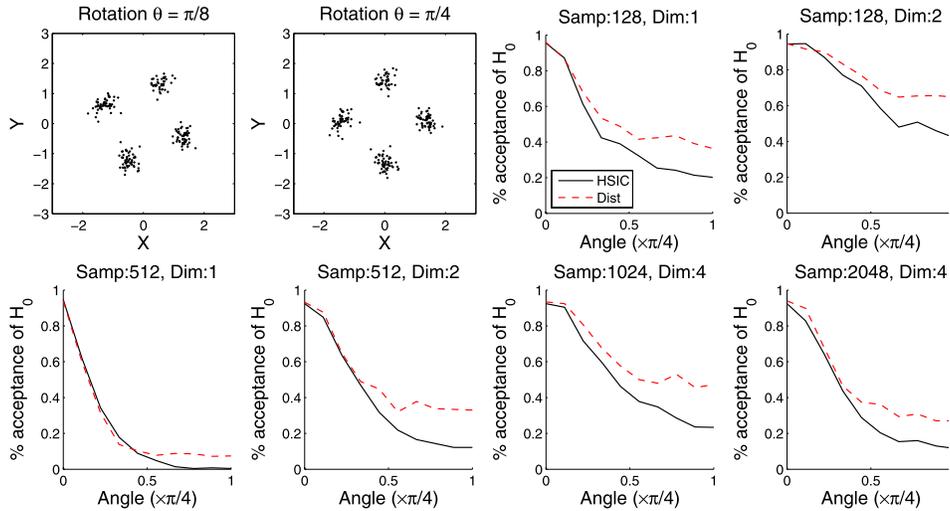}

\caption{\textup{Top left plots:} Example data set for $p=q=1$, $n=200$,
and rotation angles $\theta=\pi/8$ (left) and $\theta=\pi/4$ (right).
In this case, both sources are mixtures of two Gaussians [source \textup{(g)} of Bach and Jordan
(\protect\citeyear{BacJor02}), Figure 5]. We remark that the random variables
appear ``more dependent'' as the angle $\theta$ increases, although
their correlation is always zero. \textup{Remaining plots:} Rate of
acceptance of $\mathcal{H}_0$ for the \textup{Dist} and \textup{HSIC} tests. ``Samp''
is the number $m$ of samples, and ``dim'' is the dimension $d$ of $x$
and $y$.}
\label{fig:ISAbench}
\end{figure*}

In comparing the independence tests $\mathcal{V}_n^2$ (henceforth
denoted \textit{Dist}) and HSIC, we used an artificial benchmark proposed
by \citet{GreFukTeoSonetal08}.
We tested the independence
in two, four, and eight dimensions (i.e., $p\in{1,2,4}$ and $p=q=:d$).
We reproduce here the data description of Gretton \textit{et al.}
for ease of reference. First, we
generated $n$ samples of two independent univariate random variables,
each drawn
at random from the ICA benchmark densities
of \citeauthor{BacJor02} [(\citeyear{BacJor02}), Figure 5]: these
included super-Gaussian, sub-Gaussian,
multimodal, and unimodal distributions, with the common
property of zero mean and unit variance.
Second, we mixed these random
variables using a rotation matrix parametrized by an angle $\theta$,
varying from $0$ to $\pi/4$ (a zero angle meant the data were
independent, while dependence became easier to detect as the angle
increased to $\pi/4$; see the two plots in Figure
\ref{fig:ISAbench}). Third, in the cases $d=2$ and $d=4$,
independent Gaussian noise of zero mean and unit variance was used
to fill the remaining dimensions, and the resulting
vectors were multiplied by
independent random two- or four-dimensional orthogonal matrices, to obtain
random vectors $X$ and $Y$ dependent across all observed dimensions.
The resulting random variables were dependent but uncorrelated.
We investigated sample sizes
$n=128,512, 1024,$ and $2048$.
In estimating the the test threshold (i.e., the $1-\alpha$ quantile
of the HSIC and \textit{Dist} null distributions), we randomly
permuted the
$Y$ sample ordering $200$ times, and used the appropriate
quantile of the resulting histogram of values.
The kernel bandwidths for HSIC were set to the
median distance between samples of the respective variables.\footnote{A
Matlab implementation
of the HSIC test, including the kernel bandwidth selection step, may
be downloaded from
\url{http://www.kyb.mpg.de/bs/people/arthur/indep.htm}.
The software also includes a faster Gamma approximation to the null
distribution.}
Note that a more sophisticated
but computationally costly approach to bandwidth selection is described
by \citet{FukGreSunSch08},
which involves matching the closed-form expression for the variance
of HSIC with an estimate obtained by data shuffling.

Results are plotted in Figure \ref{fig:ISAbench} (average over 500
independent generations of the data). The $y$-intercept on these
plots corresponds to the acceptance rate of the null hypothesis
$\mathcal{H}
_0$ of independence,
or $1-{}$(Type  I  error), and should be close to the design
parameter of $1-\alpha= 0.95$. Elsewhere, the plots indicate
acceptance of $\mathcal{H}_0$ where the alternative hypothesis
$\mathcal{H}_1$ of
dependence holds,
that is, the Type II error.

We observe dependence becomes easier to detect as $\theta$
increases from 0 to $\pi/4$, when $n$ increases, and when $d$ decreases.
HSIC does as well as or better than \textit{Dist}
in all experiments, with a particular advantage at low sample sizes.
In this respect, it appears that the additional
smoothing employed by the RKHS approach
has made the associated independence test more robust.
Earlier experiments by \citet{GreFukTeoSonetal08} indicate that both
HSIC and \textit{Dist}
outperform the power-divergence statistic of \citet{ReaCre88} on these
data. This is unsurprising, since, for higher dimensions, a
space partitioning approach results in too few samples per bin.


\section*{Acknowledgments}

We would like to acknowledge Bernhard Sch\"{o}lkopf and Alexander Smola
for their collaboration on
several of the works referenced in this discussion.

\printaddresses

\end{document}